\documentclass[reprint,11pt,onecolumn]{revtex4-2}
\usepackage[utf8]{inputenc}
\usepackage{braket}
\usepackage{graphicx}
\usepackage{lipsum}
\usepackage{mathtools}
\usepackage{standalone}
\usepackage{cuted}
\usepackage{xcolor}
\usepackage{bm}% bold math
\usepackage[colorlinks,allcolors=blue]{hyperref}% add hypertext capabilities
\usepackage[mathlines]{lineno}% Enable numbering of text and display math
\usepackage{wrapfig}
\usepackage{siunitx}
\DeclareSIUnit\photon{photon}
\usepackage{dcolumn}
\usepackage{setspace}
\usepackage{booktabs}
\usepackage{tabularx} 
\usepackage{tikz}
\usetikzlibrary{fit}
\usepackage{pgfplots} %axis and plots for tikz
\usepgfplotslibrary{fillbetween} %shading on graph
\definecolor{tensorblue}{rgb}{0.8,0.8,1}
\tikzstyle{tensor}=[rectangle, draw=black, fill=tensorblue, thick, minimum size = 6mm]

\pgfplotsset{compat=1.18}
\DeclareMathOperator{\Tr}{Tr} %making a trace shortcut
\usepackage{bbold}
\tikzstyle{mynode}=[thick,draw=blue,fill=blue!20,minimum size=22] %formatting for figures

\newcommand\kb[2]{%
  {\ket{#1}\!\!\bra{#2}}
}

\usetikzlibrary{quantikz}
\usepackage[normalem]{ulem}

\date{\today}

\begin{document}
\begin{singlespace}

\title{Universal Quantum Computing with Field-Mediated Unruh--DeWitt Qubits}

\author{Eric W. Aspling}
\affiliation{Department of Physics, Applied Physics, and Astronomy, Binghamton University, Binghamton, NY 13902}
\author{Michael J. Lawler}
\affiliation{Department of Physics, Applied Physics, and Astronomy, Binghamton University, Binghamton, NY 13902}
\affiliation{Department of Physics, Cornell University, Ithaca, NY 14853}
\affiliation{Department of Physics, Harvard University, Cambridge, MA 02138}

\begin{abstract}
A set of universal quantum gates is a vital part of the theory of quantum computing, but is absent in the developing theory of Relativistic Quantum Information (RQI). Yet, the Unruh--DeWitt (UDW) detector formalism can be elevated to unitary gates between qubits and quantum fields and has allowed RQI applications in quantum Shannon theory, such as mutual information, coherent information, and quantum capacity in field-mediated quantum channels. Recently, experimental realizations of UDW-style qubits have been proposed in two-dimensional quantum materials, but their value as a quantum technology, including quantum communication and computation, is not yet clear, especially since fields introduce many avenues for decoherence. We introduce controlled-unitary UDW logic gates between qubit and field that are comparable to the two-qubit CNOT gate. We then extend this formalism to demonstrate Quantum State Transfer (QST) (two CNOT gates) and SWAP (three CNOT gates) channels. We illustrate the performance of these quantum operation gates with the diamond distance, a measure of  distinguishability between quantum channels.  Distinguishability measures like diamond distance allow for a rigorous comparison between field-mediated transduction through UDW detectors and local quantum mechanical operations and so quantify the performance of UDW detectors in quantum technological applications. Using the controlled-unitary qubit-field interactions we define an exact form of the CNOT gate. With this technique we also define quantum field-mediated single qubit operations associated with the Hadamard $H$, the $S$, and $T$ gates. Thus, UDW detectors in simple settings enable a collection of gates known to provide universal quantum computing. 
\end{abstract}
\maketitle

\section{Introduction}\label{Introduction}
In the landscape of quantum networking, the importance of qubit-field quantum transduction cannot be overstated. The integration of Quantum Field Theory (QFT) into quantum computing, utilizing Relativistic Quantum Information (RQI) through controlled Unruh-DeWitt (UDW) detector interactions, has achieved significant theoretical progress.\cite{Tjoa2021What,Simidzija2020Transmission, MartinMartinez2015Causality,MartinMartinez2013Sustainable,Aspling2023High}. 
Yet we could have fast all-to-all connected quantum computers\cite{Aspling2023Design} if researchers can develop RQI for condensed matter and atomic, molecular, and optical(AMO) systems via qubit-field transduction. %However, progress in developing RQI for condensed matter and atomic, molecular, and optical systems  via qubit-field transduction has been limited. 
We could also understand relativistic effects arising in complex interactions between qubits and fields in long-range quantum communication theories.
%Understanding the complex interactions between qubits and fields is essential for long-range quantum communication theories, accounting for relativistic effects. 
Hence, we need a cross-disciplinary model that brings RQI to condensed matter and AMO systems, facilitates noise calculations, and clarifies the complexity of field-mediated communication needed for useful applications.

At its conception, the UDW detector described an accelerating two-state system coupled to a thermal bath and sensitive to radiation \cite{Unruh1976Notes,DeWitt1980quantum}. If the detector interacted with radiation, it would find itself in the excited state with energy equivalent to that which is hidden behind the killing horizon in the accelerated frame of reference \cite{carroll2019spacetime}. Underneath the relativistic and high-energy framework lies a Hamiltonian that describes a two-state detector coupled to a quantum field through smearing and switching functions. Regardless of detector acceleration, this setup allows for a successful theoretical approach to coupling qubits to fields to calculate quantum information quantities in relativistic settings. 

Long-range communication of photons in non-inertial frames of reference, such as the gravitational well of the Earth, indicate the need for relativistic theories to explore relativistic effects \cite{Rideout2012Fundamental,ralph2007gravitationally}. These effects impact a system's ability to communicate quantum information specifically in the regime of ground-to-satellite communication \cite{PerdiguesArmengol2008Quantum,Rarity2002Ground,Xin2011Chinese,Ke2022Optical,lou2015atmosphere,Kehayas2017European,Sidhu2021Advances}. Understanding and mitigating these relativistic effects is crucial for advancing the reliability and efficiency of quantum communication systems in practical applications.

There are other, more down-to-earth level, applications of the UDW detector model for Quantum Information Science (QIS) purposes. One is in the arena of photonics, where there has been a large effort in 1-D quantum state transport through waveguides \cite{Mele2023Optical,Mele2022Restoring,Xavier2011Active,Storz2023Loophole,Kurpiers2018Deterministic,Magnard2020Microwave}. On the other hand, lesser-known fermionic application provides a means to probe quantum communication in quantum materials \cite{Aspling2023Design} utilizing QFT applications of condensed matter systems. Further extensions to these models include QIS in Valleytronics\cite{Vitale2018Vallytronics}, Mo\`ire pattern materials\cite{tran2019evidence}, and doped Transition Metal Dichalcogenides\cite{tsai2022antisite,Lee2022Spin,OrtizJimenez2021Light}, all of which classify as a UDW quantum computer when coupled to spin qubits, modeled as Kondo impurities. For this reason, developing field-mediated quantum logic gates will enable cross-discipline discussion of UDW quantum computers as they become implemented in experimental settings. 

Quantum systems are highly prone to noise and error\cite{Preskill2018Quantum} and often require error-correcting codes \cite{Nielsen2010Quantum}. Field-mediated quantum communication will offer novel solutions to this problem, but also generate a new series of errors involving qubit-field interactions. Relativistic effects, long-range information dissipation, and engineering limitations all generate noise and without a generalized structure to examine these interactions numerically in the landscape of quantum computing, these noisy interactions remain mysterious.  

In this manuscript, we propose a solution to reduce the complexity of all these applications of UDW detectors: build a universal set of UDW quantum logic gates. We achieve this in a pedogogical way by firstly demonstrating canonical quantum logic gates through the utilization of projective operators. We then outline how to achieve logic operations with coherent states presented in UDW formalism. With this formalism in place, we introduce several quantum logic gates, paying close attention to the canonical quantum state transfer channel, as it is usually presented in the form of a combination of CNOT gates. We then show that the ``encode" and ``decode" gates of the UDW model, which are introduced in \cite{Simidzija2020Transmission} and later extended in \cite{Aspling2023High}, demonstrate the quantum state transfer channel with UDW detectors. 

Through the processes presented in this manuscript, we introduce key UDW quantum logic gates that are exact to a set of universal quantum computing gates. These gates include field-mediated gates which are equivalent to two-qubit CNOT gates as well as field-mediated single gate operations; T-gate, S-gate, and Hadamard. With this generalized structure in place we close this letter by introducing some open problems of noise whose solutions will help realize qubit-field transduction in the aforementioned systems.

\section{Background}\label{Background}
Before jumping into the formalism of qubit-field gates using UDW detectors, here we outline the approach we will take using the familiar setting of standard qubits in place of quantum fields. We introduce certain constructs that may appear unnecessary or unfamiliar (e.g. summing over projection operators), but this pedagogy mirrors the UDW formalism by utilizing these same operators along with field operators.

\subsection{Projectors and Quantum Logic Gates}
\begin{wraptable}{r}{4.5cm}
\vspace{-5em}
\begin{center}
    \begin{tabular}{|c|c|}
    \hline
    Input& Output\\
    \hline
    $\ket{0}$&$\ket{1}$\\
    $\ket{1}$&$\ket{0}$ \\
    % Add more rows as needed
    \hline
    \end{tabular}
    \caption{The NOT gate changes the value of the qubit; i.e., $\ket{0}$ becomes $\ket{1}$ and vice versa}
    \label{NOT truth table}
\end{center}
\begin{center}
    \begin{tabular}{|c|c|}
        \hline
        Input&Output\\
        \hline
        $\ket{0}$&$\ket{0}$\\
        $\ket{1}$&$-\ket{1}$ \\
        % Add more rows as needed
        \hline
    \end{tabular}
    \caption{Pauli-Z truth table negates the state $\ket{1}$ while leaving the $\ket{0}$ state untouched.}
    \label{Pauli-Z truth table}
\end{center}
\begin{center}
    \begin{tabular}{|c|c|}
        \hline
        Input&Output\\
        \hline
        $\ket{0}$&$-i\ket{1}$\\
        $\ket{1}$&$i\ket{0}$ \\
        % Add more rows as needed
        \hline
    \end{tabular}
    \caption{The Pauli-Y truth table demonstrates the rotation onto the complex plane. }
    \label{Pauli-Y truth table}    
\end{center}
\vspace{-5em}
\end{wraptable}
Behind every quantum logic gate is a series of Pauli and projection operators in computational basis states. A common example of this is the quantum ``NOT" gate. Commonly denoted as $\hat{X}$, the quantum NOT gate, also known as the Pauli-X gate, has the mathematical forms
\begin{equation}
    \hat{X}= \hat{\sigma}^x = \sum_{\mu\in \pm}\mu\hat{P}^{\mu}_X=\kb{+}{+}-\kb{-}{-} =\kb{1}{0}+\kb{0}{1}=
    \begin{pmatrix}
    0 & 1\\ 
    1 & 0
    \end{pmatrix} .
\end{equation}
Here we have defined the projector $\hat{P}^{\mu}_X \equiv \ket{\mu_x}\bra{\mu_x}$ with $\ket{\mu_x}$ being the basis states in the x-basis. Similar projectors can be constructed for the z- and y-basis. 

A truth table is a stylistic representation of input and output states that displays the changes that are undergone through a given process. We can generate a truth table for the NOT gate that has the following form.

The other Pauli matrices have a similar formalism 

\begin{flalign}
    \hat{Z}&= \hat{\sigma}^z =  \sum_{\mu\in \pm}\mu\hat{P}^{\mu}_Z= \kb{0}{0}-\kb{1}{1}=
    \begin{pmatrix}
    1 & 0\\ 
    0 & -1
    \end{pmatrix} \\
    \hat{Y}&= \hat{\sigma}^y = \sum_{\mu\in \pm}\mu\hat{P}^{\mu}_Y = -i(\kb{1}{0}-\kb{0}{1})=
    \begin{pmatrix}
    0 & -i\\ 
    i & 0
    \end{pmatrix},
\end{flalign}
with corresponding truth tables given in Tables~\ref{Pauli-Z truth table} and \ref{Pauli-Y truth table}.

With these Pauli and projection operators we can now examine controlled unitary gates. Controlled gates act on multiple qubits, where some qubits are the ``controls" and the other qubits act as the ``targets". For reasons to be made clear, we will utilize the example of the Controlled NOT (CNOT) gate with projector form
\begin{equation}\label{CNOTX}
    \mathrm{CNOT}(1,2) = \sum_{\mu \in \pm}\hat{P}^{\mu}_z \otimes \hat{X}^{\frac{-\mu+1}{2}} = \kb{0}{0} \otimes \mathbb{1} + \kb{1}{1}\otimes \hat{X}.
\end{equation}

To construct a reverse CNOT gate we can interchange the tensored values such as
\begin{equation}
    \mathrm{CNOT}(2,1) = \sum_{\mu \in \pm}\hat{X}^{\frac{-\mu+1}{2}} \otimes \hat{P}^{\mu}_z
\end{equation}
but the CNOT gate also has the alternative form 
\begin{equation}\label{CNOTZ}
\mathrm{CNOT}(2,1) = \sum_{\mu \in \pm}\hat{P}^{\mu}_x \otimes \hat{Z}^{\frac{-\mu+1}{2}} = \kb{+}{+} \otimes \mathbb{1} + \kb{-}{-}\otimes \hat{Z}.
\end{equation}
This alternative form will be helpful when considering field-mediated logic gates. These two CNOT gates are demonstrated with the following truth tables;
\begin{table}[ht]
\def\arraystretch{1.2}
    \begin{minipage}{0.32\hsize}
    \begin{tabular}{|c|c|c|c|}
        \hline
        \multicolumn{2} {|c} {Input} & \multicolumn{2} {c|} {Output}\\
        \hline
        Control&Target&Control&Target\\
        $\ket{0}$&$\ket{0}$&$\ket{0}$&$\ket{0}$\\
        $\ket{0}$&$\ket{1}$&$\ket{0}$&$\ket{1}$\\
        $\ket{1}$&$\ket{0}$&$\ket{1}$&$\ket{1}$\\
        $\ket{1}$&$\ket{1}$&$\ket{1}$&$\ket{0}$\\
        % Add more rows as needed
        \hline
    \end{tabular}
    \caption{A CNOT(1,2) will flip the entry of the target gate when the control gate is the $\ket{1}$ state.}
    \label{CNOT12 truth table}
    \end{minipage}
    \hspace{1cm}
    \begin{minipage}{0.32\hsize}
    \begin{tabular}{|c|c|c|c|}
        \hline
       \multicolumn{2} {|c} {Input} & \multicolumn{2} {c|} {Output}\\
        \hline
        Target&Control&Target&Control\\
        $\ket{0}$&$\ket{0}$&$\ket{0}$&$\ket{0}$\\
        $\ket{0}$&$\ket{1}$&$\ket{1}$&$\ket{1}$\\
        $\ket{1}$&$\ket{0}$&$\ket{1}$&$\ket{0}$\\
        $\ket{1}$&$\ket{1}$&$\ket{0}$&$\ket{1}$\\
        % Add more rows as needed
        \hline
    \end{tabular}
    \caption{The CNOT(2,1) gate switched the roles of the qubits in the system.}
    \label{CNOT21 truth table}
    \end{minipage}
\end{table}

It may not be clear at first glance that these ``sum of projectors" is unitary, but one can verify this though the identity $\hat{U}\hat{U}^{\dagger} = \hat{I}$ applied to our operator forms such that
\begin{flalign}
     \sum_{\mu\mu' \in \pm}(\hat{P}^{\mu}_z \otimes \hat{X}^{\frac{-\mu+1}{2}})(\hat{P}^{\mu'}_z \otimes \hat{X}^{\frac{-\mu'+1}{2}})^{\dagger} =& \sum_{\mu\mu' \in \pm}\hat{P}^{\mu}_z\hat{P}^{\mu'}_z \otimes \hat{X}^{\frac{-\mu+1}{2}}\hat{X}^{\frac{-\mu'+1}{2}}\\ =& \sum_{\mu \in \pm}\hat{P}^{\mu}_z \otimes \hat{X}^{-\mu+1} \\=&(\kb{+}{+} \otimes \hat{I})+(\kb{-}{-}\otimes \hat{I}) = \hat{I}
\end{flalign}
where the second equivalence follows from $\hat{P}^{\mu}_i\hat{P}^{\mu'}_i = \delta_{\mu \mu`}\hat{P}^{\mu}_i\hat{P}^{\mu'}_i = (\hat{P}^{\mu}_i)^2 = \hat{P}^{\mu}_i$ and $i\in\{x,y,z\}$.
\subsection{Quantum Computing Operations from Logic Gates}
\subsubsection{Quantum State Transfer (QST)}
\begin{wrapfigure}{r}{9.5cm}
\vspace{-2em}
\centering
\includegraphics{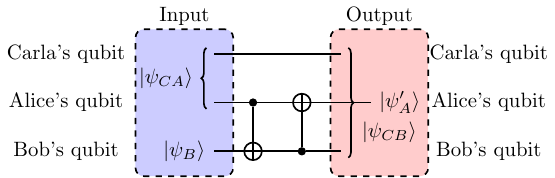}
\caption{Two CNOT gates form a State Transfer Gate}\label{ST Channel Circuit}
% \vspace{-5em}
\end{wrapfigure}

It is well-known that some fundamental quantum processes can be achieved through the combinations of single and multiple qubit quantum logic gates. One such example of this process is the QST channel. A canonical QST passes quantum information uni-directionally through a quantum circuit. For transferring one-qubit quantum states, QST is accomplished with two CNOT gates and has the projector form of 
\begin{equation}\label{QST}
    \mathrm{QST} = \mathrm{CNOT}(1,2)\mathrm{CNOT}(2,1) =\sum_{\mu,\mu' \in \pm} \hat{P}^{\mu}_z \hat{P}^{\mu'}_x\otimes \hat{X}^{\frac{-\mu+1}{2}}\hat{Z}^{\frac{-\mu'+1}{2}}
\end{equation}
and demonstrated in the circuit diagram of Fig.~\ref{ST Channel Circuit}.
In matrix form this operation between Alice and Bob can be expressed as,
\begin{equation}
\mathrm{QST}\equiv
\begin{pmatrix}
1 & 0 & 0 & 0\\
0 & 0 & 1 & 0\\
0 & 0 & 0 & 1\\
0 & 1 & 0 & 0
\end{pmatrix} 
\end{equation}
and as shown in Fig.~\ref{ST Channel Circuit} it will take the state $\ket{\psi_{CA}} \otimes \ket{\psi_B}$ to $\ket{\psi'_A} \otimes\ket{\psi_{CB}}$. Notice that the output state $\ket{\psi'_A}$ may not have the same form as $\ket{\psi_B}$. This result can be made more evident by adding a reference state $\ket{\psi_D}$ to Bob's qubit as in Fig.~\ref{2 ref. ST Channel Circuit}. 

\subsubsection{SWAP gate}\label{section SWAP gate}
To create a proper SWAP a third CNOT gate is needed. A canonical form of the SWAP gate can be seen using three CNOT gates in the following forms 
\begin{align}\label{swap equation}
    \mathrm{SWAP}&= \mathrm{CNOT}(2,1)\mathrm{CNOT}(1,2)\mathrm{CNOT}(2,1)\\
    &=\sum_{\mu,\mu',\mu'' \in \pm} \hat{P}^{\mu''}_x \hat{P}^{\mu}_z \hat{P}^{\mu'}_x\otimes \hat{Z}^{\frac{-\mu''+1}{2}}\hat{X}^{\frac{-\mu+1}{2}}\hat{Z}^{\frac{-\mu'+1}{2}}\\
    &=(\hat{P}^+_x\otimes\mathbb{1}+\hat{P}^-_x\otimes\hat{Z})(\hat{P}^+_z\otimes \mathbb{1} + \hat{P}^-_z\otimes \hat{X})(\hat{P}^+_x\otimes\mathbb{1}+\hat{P}^-_x\otimes\hat{Z})
\end{align}

\begin{figure}[ht]
\centering
\includegraphics{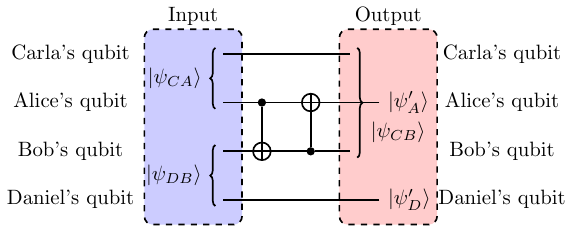}
\caption{Two CNOT gates form a State Transfer Gate as shown with reference qubits.}\label{2 ref. ST Channel Circuit}
\end{figure}

\begin{figure}
\centering
\includegraphics{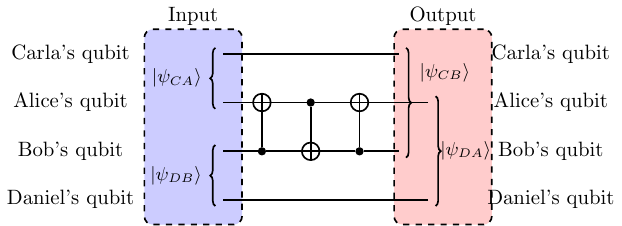}
\caption{Proper SWAP circuit between qubits.}\label{2 ref. Swap Channel Circuit}
\end{figure}

Figure~\ref{2 ref. Swap Channel Circuit} demonstrates the entanglement transfer between Alice and Bob's qubits, which, unlike the QST circuit in Fig.~\ref{2 ref. ST Channel Circuit}, exchanges the quantum information between these two qubits bidirectionally. 

\subsection{Intermediate Qubit}\label{Intermediate Qubit}
\begin{figure}[ht]
\centering
\includegraphics{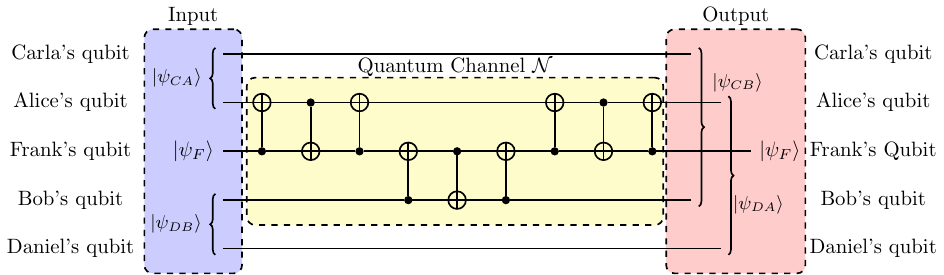}
\caption{Information is swapped between Alice and Bob through Frank.}\label{2 ref. intermediary Swap Channel Circuit}
\end{figure}

A quantum channel is a completely positive trace-preserving mapping, that maps an initial state to some final state through a series of operators. To bring us closer to the field-mediated channels, we next demonstrate how these unitary gates can be implemented via qubit-mediated quantum channels. The circuit diagram in Fig.~\ref{2 ref. intermediary Swap Channel Circuit} demonstrates a canonical SWAP interaction between two qubits expressed in terms of a sequence of 9 CNOT gates \emph{mediated by Frank's third qubit}.
It is through this quantum channel $\mathcal{N} (\hat{\rho}_{AFB})$, with initial state $\hat{\rho}_{AFB}=\hat{\rho}_{A,0}\otimes\kb{\psi_F}{\psi_F} \otimes \hat{\rho}_{B,0}$ where  $\rho_{A,0} = \text{Tr}_C|\psi_{CA}\rangle\langle\psi_{CA}|$ and $\rho_{B,0} = \text{Tr}_D|\psi_{BD}\rangle\langle\psi_{BD}|$ (mixed states generated by tracing over Carla and Bob), that we exchange entanglement freely between Alice and Bob. To construct this channel we can assign our SWAP gates to unitary operators. Let's call $\hat{U}_{AF}$ the unitary operator that swaps between Alice and Frank and $\hat{U}_{FB}$ for Frank and Bob. Our channel from Alice to Bob is therefore
\begin{equation}
    \mathcal{N}_{A\rightarrow B}(\hat{\rho}_{AFB}) = \Tr_{AF}[\hat{U}_{AF} \hat{U}_{FB} \hat{U}_{AF}( \hat{\rho}_{A,0}\otimes\kb{\psi_F}{\psi_F} \otimes \hat{\rho}_{B,0}) \hat{U}^{\dagger}_{AF} \hat{U}^{\dagger}_{FB} \hat{U}^{\dagger}_{AF} ].
\end{equation}
The channel from Bob to Alice, $\mathcal{N}_{B \rightarrow A}(\hat{\rho}_{AFB})$ can be seen by tracing over Bob and Frank. Though quantum channels are in general non-unitary, this case is special, producing a unitary operation on Alice and Bob's Hilbert space as expected.

Quantum information channels are very powerful tools to evaluate rates of information transfer. For a deep dive into the properties of quantum information channels See Refs.~\cite{Aspling2022Introduction,Wilde2011From}. One such quantitative measure of the ability for Bob and Alice to communicate with quantum information, can be seen with the quantum channel capacity. Quantum capacity utilizes the maximum coherent information attained from maximizing over the arrangements of input states of any given channel. For this work, we assume we are working with the initial state that yields that maximum coherent information, allowing the equivalence
\begin{equation}
   Q(\mathcal{N}_{A\rightarrow B}(\hat{\rho}_{AFB}))=I_c(\mathbb{1}_C \rangle \mathcal{N}_{A\rightarrow B}(\hat{\rho}_{AFB})) = S(\mathcal{N}_{A\rightarrow B}(\hat{\rho}_{AFB})) - S(\mathbb{1}_C \otimes \mathcal{N}_{A\rightarrow B}(\hat{\rho}_{AFB})) 
\end{equation}
where $Q(\mathcal{N}_{A\rightarrow B}(\hat{\rho}_{AFB}))$ is the quantum capacity of the channel $\mathcal{N}_{A\rightarrow B}(\hat{\rho}_{AFB})$, $I_c(\mathbb{1}_C \rangle \mathcal{N}_{A\rightarrow B}(\hat{\rho}_{AFB}))$ is the coherent information of the channel with respect to Carla, and $S(\rho) = -\rho \Tr \rho$ is the Von Neumann entropy. 
\section{Introducing Fields into the Channels}
\subsection{Coherent Basis States of the Field}\label{Coherent Basis States of the Field}
We now upgrade Frank from a two-state qubit to an entire quantum field. To make progress on this complex setting, we will work with the set of coherent states that spans the infinite Hilbert space $\mathcal{H}_\varphi$ of the quantum field
\begin{equation}\label{field coherent state definition 1st}
    \ket{\{\beta({\bf k})\}} = \exp{\left(\int d^n \textbf{k}\left[\beta({\bf k})\hat{a}_{\textbf{k}}^{\dagger}-\beta({\bf k})^*\hat{a}_{\textbf{k}}\right]\right)}|0\rangle
\end{equation}
where $|0\rangle$ is the ground state of the field satisfying $\hat{a}({\bf k})|0\rangle = 0\, \forall\, {\bf k}$. Within this Hilbert space, there exists a \emph{particular} two-dimensional subspace spanned by $\ket{\pm\alpha}$ of $\mathcal{H}_{\varphi}$ that we can utilize for encoding and decoding quantum information  
\begin{equation}\label{field coherent state definition ch.6}
    \ket{\pm \alpha} \equiv \hat{D}(\alpha(k))\ket{0} = \exp{\left(\int d^n \textbf{k}\left[\alpha(k)\hat{a}_{\textbf{K}}^{\dagger}-\alpha(k)^*\hat{a}_{\textbf{K}}\right]\right)}\ket{0}
\end{equation}
where $\hat{D}(\alpha(k))$ is the unitary displacement operator, and subsequently sets the normalization as $|\braket{\pm\alpha|\pm\alpha}| =1$. The function $\alpha({\bf k})$ that defines our special two-dimensional subspace is a function of parameters that are chosen such that $\ket{\pm \alpha}$ is nearly orthogonal, and therefore \emph{sets up a qubit encoding}. We describe these parameters in detail in the following section.

To realize the displacement function $\hat{D}(\alpha(k))$, we utilize the 1-D scalar field $\hat{\varphi}(x)$ and it's associated conjugate momentum $\hat{\Pi}(x)$ which are traditionally given in the Heisenberg picture as
\begin{flalign}
     \hat{\varphi}(x) &= \int\frac{dk}{2\pi} \sqrt{\frac{v}{2\omega(k)}}[\hat{a}(k)e^{ikx}+\hat{a}^{\dagger}(k)e^{-ikx}] \label{scalar field}\\
     \hat{\Pi}(x) &=\int\frac{dk}{2\pi} \sqrt{\frac{\omega(k)}{2v}}[-i\hat{a}(k)e^{ikx}+i\hat{a}^{\dagger}(k)e^{-ikx}].\label{conjugate momentum}
\end{flalign}
These field operators allow us to generate coherent states with amplitude
\begin{equation}\label{real space coherent amplitude}
    \alpha_{\varphi}(x) = \sqrt{\frac{v}{2\omega(k)}}e^{-ikx}.
\end{equation} 
For convenience we have introduced subscripts to differentiate between coherent amplitudes of the  field and conjugate momentum observables. A Fourier transform following the usual prescription
\begin{equation}
f(k) \coloneqq \frac{1}{\sqrt{(2\pi)}}\int dx f(x) e^{ikx}
\end{equation}
allows us to recover our function $\alpha(k)$ (in one-dimension) but to make these states near-orthogonal we turn to the UDW formalism detailed in Ref. \cite{Simidzija2020Transmission}. 

\subsection{Encoding and Decoding and the UDW QST}
Since quantum fields acting on the vacuum generate coherent states, our computational basis of quantum fields, we can utilize the field operators above to generate unitary operations which will encode and decode quantum information onto and off of our fields via coherent states. As we will show, the effectiveness of this encoding and decoding relies on constraints that enforce an orthognality condition on our coherent states. With these constraints in place we can begin generating field-mediated processes similar to the ones outlined in Sec.~\ref{Background}
\subsubsection{Necessary Constraints}\label{Necessary Constraints}
For our states $\ket{\pm \alpha}$ to be near orthogonal certain constraints need to be implemented. The hallmark of the UDW model is the smearing and switching functions in the standard UDW interaction Hamiltonian
\begin{equation}\label{UDW model}
    \hat{\mathrm{H}}_{int}(t) =  J_{\varphi}\chi(t) \int_{\mathbb{R}} dk \ \tilde{F}(k) \hat{\mu}(t) \otimes \hat{\varphi}(k,t)
\end{equation}
with coupling $J_{\varphi}$, smearing function $\tilde{F}(k)$, switching function $\chi(t)$, and two-state magnetic moment $\hat{\mu}(t)$. For simplicity we only evaluate the interaction block Eq.~\eqref{UDW model}. However in Sec.~\ref{Open Problems of Determining Noise}, we discuss how other blocks of the Hamiltonian can be used to calculate noise with this prescription.

A delta-like switching function allows for the promotion of our time-dependent Hamiltonian to unitary gates \cite{Simidzija2020Transmission,Aspling2023High}. Following the standard UDW prescription we absorb the coupling and smearing function into smeared-out field observables
\begin{flalign}
    \hat{\varphi}(F) &\coloneqq \lambda_{\varphi} \int dk F(k) \hat{\varphi}(k,t)\label{redefined scalar}\\
    \hat{\Pi}(F) &\coloneqq \lambda_{\Pi} \int dk F(k) \hat{\Pi}(k,t)\label{redefined conjugate momentum}.
\end{flalign}
For the sake of brevity, we will use these smeared versions of the field observables when written out in unitary gate form, and therefore suppress the function $F$ from our notation via $\hat{\varphi}(F)\equiv\hat{\varphi}$.

We summarize the constraints given in Ref.~\cite{Simidzija2020Transmission} here; in the regime of strong coupling we find $|\braket{+\alpha|-\alpha}| \approx 0$, an atypical result of the nonorthogonal coherent state basis. This result is made evident by constraining properties of the UDW model through the inner product of these coherent states given by
\begin{equation}
    |\braket{+\alpha|-\alpha}| = \exp \left( - (J_{\varphi})^2 \int \frac{dk}{2\omega_k}|\tilde{F}(k)|^2\right) \label{innerproduct of coherent states}
\end{equation}
 which follows from the identity 
\begin{equation}\label{coherent state orthogonality identity}
    \braket{\beta|\alpha} = \exp\left(-\frac{1}{2}|\alpha|^2-\frac{1}{2}|\beta|^2 + \beta^*\alpha \right).
\end{equation}
It is clear from Eq.~\eqref{innerproduct of coherent states} that as the coupling $J_{\varphi}$ increases, the states become increasingly orthogonal. 
\subsubsection{UDW QST}
Elevating the Hamiltonian in Eq.~\eqref{UDW model} to include both the field and scalar momentum 
\begin{equation}\label{UDW model ch.6}
    \hat{\mathrm{H}}^{QST}_{\mathrm{int}}(t) =  J_{\varphi}\chi(t) \int_{\mathbb{R}} dx \ F(x) \hat{\mu}(t) \otimes (\hat{\varphi}(x,t)+\hat{\Pi}(x,t)).
\end{equation}
and using a delta like $\chi(t)$, we can use Eq.~\eqref{UDW model ch.6} to formulate unitaries that carry out the encoding and decoding demonstrated in the circuit diagram in Fig.~\ref{UDWQC Channel with starburst}
\begin{equation}\label{unitary as sum of projectors}
    \hat{U}^{QST}_{\mathrm{QF}} = \sum_{\mu,\mu'\in \pm} \hat{P}^{\mu}_x\hat{P}^{\mu'}_z \otimes e^{i\mu\hat{\Pi}}e^{i\mu'\hat{\varphi}} = (\hat{P}^-_x\otimes e^{-i\hat{\Pi}} + \hat{P}^+_x \otimes e^{i\hat{\Pi}})(\hat{P}^+_z \otimes e^{i\hat{\varphi}} + \hat{P}^-_z \otimes e^{-i\hat{\varphi}})
\end{equation}
which looks remarkably similar to Eq.~\eqref{QST}. This unitary behaves like two controlled unitaries, as we will see at the end of this section.

In order to remain in the two-dimensional subspace of $\mathcal{H}_{\varphi}$ we follow the prescription of Ref.~\cite{Simidzija2020Transmission} and enable the constraint
\begin{equation}
\hat{\Pi} \ket{\pm \alpha} \approx \pm \gamma \ket{\pm \alpha}
\end{equation}
where the value of $\gamma$ is set by 
\begin{equation}\label{capital gamma constraint}
    \gamma \coloneqq \lambda_{\Pi} \lambda_{\varphi} \int dk \, |\tilde{F}_{\nu}(k)|^2 = \frac{\pi}{4}\, \mathrm{mod}\, 2\pi
\end{equation}
a consequence of the restriction
\begin{equation}
    \left( \lambda_{\varphi} \int dk \, |\tilde{F}_{\nu}(k)|^2 \right)^2 \gg \frac{1}{2} \int dk \, \omega(k)\, |\tilde{F}_{\nu}(k)|^2. 
\end{equation} 

Constraining the parameters in this way allows the operator $e^{i\hat{\Pi}}$ to act on the state and introduce a phase 
\begin{equation}
    e^{i\hat{\Pi}}\ket{\mu \alpha}=e^{i\mu\gamma}\ket{\mu \alpha}.
\end{equation}
Practically, we can see this through the following example. If we initialize our state as $\ket{\psi_{CA}} \otimes \ket{\psi_{\varphi}} = \ket{000}+\ket{110}$ and apply our unitary from Eq.~\eqref{unitary as sum of projectors} then our output state is given by
\begin{equation}\label{output state of QST}
    e^{\gamma}( \ket{0}_C\ket{+}_A\ket{\alpha}_{\varphi}-\ket{1}_C\ket{-}_A\ket{-\alpha}_{\varphi}) + e^{-\gamma}(\ket{0}_C\ket{-}_A\ket{\alpha}_{\varphi}+\ket{1}_C\ket{+}_A\ket{-\alpha}_{\varphi}).
\end{equation}
From this output state we can see that the qubit-field QST operation is different than the qubit-qubit QST. It is not a simple quantum state transfer from qubit to field. While qubit to field QST may be attainable through this prescription, for this work we maintain focus on field-mediated qubit communications. Despite the difference in the way these QST operations treat quantum information, we will find in Sec.~\ref{Applications in Quantum Shannon Theory: Diamond Norm} that there is an equivalence in the field-mediated QST channel and the qubit-mediated counterpart. 

\subsection{SWAP gate with fields}
Generating field-mediated SWAP gates is no trivial task, given the difference in the treatment of quantum information explained above. However, one can generate an overly simplified model, where we define projection and ``Pauli" operators of coherent state density matrices. Using these operators we can demonstrate how this simplified model can be utilized to give the same results of Sec.~\ref{section SWAP gate}. 

\subsubsection{Coherent State Projection Operators}
With the above constraints we can imagine a simplified form by constructing projection operators of coherent states without the field observable form and carry out standard quantum logic gates. Let's define these operators as
\begin{align}\label{coherent state projectors} 
    \hat{\mathrm{P}}_{+\alpha} &= \kb{+\alpha}{+\alpha} & \hat{\mathrm{P}}_{-\alpha} &= \kb{-\alpha}{-\alpha} \\ \nonumber\hat{Z}_{\alpha} &= \kb{+\alpha}{+\alpha} - \kb{-\alpha}{-\alpha} &  \hat{X}_\alpha &= \kb{-\alpha}{+\alpha} + \kb{+\alpha}{-\alpha}\\ \nonumber
    \hat{\mathrm{P}}^{\frac{\pi}{2}}_{+\alpha} &=\kb{+\alpha}{+\alpha} + \kb{-\alpha}{+\alpha} + \kb{+\alpha}{-\alpha} + \kb{-\alpha}{-\alpha} &  \\ \nonumber \hat{\mathrm{P}}^{\frac{\pi}{2}}_{-\alpha} &=\kb{+\alpha}{+\alpha} - \kb{-\alpha}{+\alpha} - \kb{+\alpha}{-\alpha} + \kb{-\alpha}{-\alpha}.
\end{align}
However, to utilize these operators, there are operational costs depending on the field initialization. Commonly, the field is initialized in the ground state, which results in an extra phase factor of $e^{-\frac{1}{2}|\alpha|^2}$ as a result of the non-orthogonality identity of coherent states given in Eq.~\eqref{coherent state orthogonality identity}. 
\subsubsection{Building Simplified Field-Mediated SWAP gates}
Ignoring physical implementation for a moment we can utilize the projectors defined in Eqs.~\eqref{coherent state projectors} which will allow for a straight-forward SWAP gate of the form
\begin{equation}
    \hat{U}^{\mathrm{SWAP}}_{\mathrm{QF}} = (\hat{P}^+_z \otimes \mathbb{1} + \hat{P}^-_z \otimes \hat{X}_{\alpha})(\hat{P}^+_x \otimes \mathbb{1} + \hat{P}^-_x \otimes \hat{Z}_{\alpha})(\hat{P}^+_z  \otimes e^{i\hat{\varphi}_{\nu}} + \hat{P}^-_z \otimes e^{-i\hat{\varphi}_{\nu}}).
\end{equation}
Alternatively, we can initialize our field in the state $\ket{+\alpha}$ and use 
 \begin{equation}
     \hat{U}^{\mathrm{SWAP}}_{\mathrm{QF}} = (\hat{P}^+_z \otimes \mathbb{1} + \hat{P}^-_z \otimes \hat{X}_{\alpha})(\hat{P}^+_x \otimes \mathbb{1} + \hat{P}^-_x \otimes \hat{Z}_{\alpha})(\hat{P}^+_z  \otimes \mathbb{1} + \hat{P}^-_z \otimes  \hat{X}_{\alpha})),
\end{equation}
which is in direct analogy with Eq.~\eqref{swap equation} of Sec.~\ref{section SWAP gate}.

While the field operator formalism for the SWAP gate is not obvious, we will see in Sec.~\ref{Forms of a single CNOT channel} that with some care, one can write down a two-qubit single CNOT with our field-mediated gates. For this reason, we can in theory, minimally write down a cumbersome and computationally exhaustive SWAP gate using three of these CNOT gates. However, it is likely that a more simplified version of this gate exists in field operator formalism, just as Eq.~\eqref{unitary as sum of projectors} exists as a simplified version of two individual field-mediated CNOTs in sequence. We have not sought such a result here since our goal is merely to find these operators in principle.

\section{Applications in Quantum Shannon Theory: Diamond Norm} \label{Applications in Quantum Shannon Theory: Diamond Norm}
The previous section outlined a prescription for quantum operations, that in a particular limit, replicate a quantum state transfer operation between qubits and a SWAP operation. Quantitative measures outlined in Sec.~\ref{Intermediate Qubit} provide numerical insight into aforementioned field-mediated operations between qubits. Coherent information and quantum capacity have been utilized to examine the communication properties of UDW channels \cite{Simidzija2020Transmission}. However, a straightforward comparison between the field-mediated UDW channels and the qubit-mediated quantum logic gate channels provides deeper insight into a system's ability to match processes. 
\begin{figure} [ht]
\centering
\includegraphics{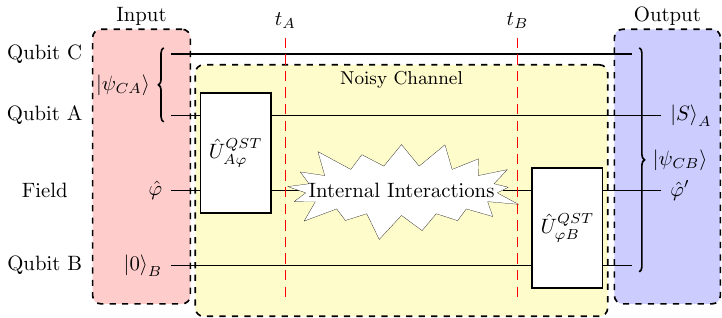}
\caption{Encoding and Decoding quantum information onto and off of the field $\hat{\varphi}$ through the QST unitaries of Eq.~\eqref{unitary as sum of projectors}.} \label{UDWQC Channel with starburst}
\end{figure}

The diamond norm can indicate the differences between two quantum channels as a measure of distinguishability \cite{Dorit1998Quantum,Benenti2010Computing,Wilde2011From}. It is well understood that the SWAP gate leads to ideal channels for transferring quantum information bilaterally between qubits. Similarly QST, in the form of two CNOT gates, provides an ideal unidirectional information transfer. So we can compare the field-mediated QST channel to the qubit-mediated QST channel utilizing the diamond distance.

\begin{figure}[ht]
\centering
\includegraphics{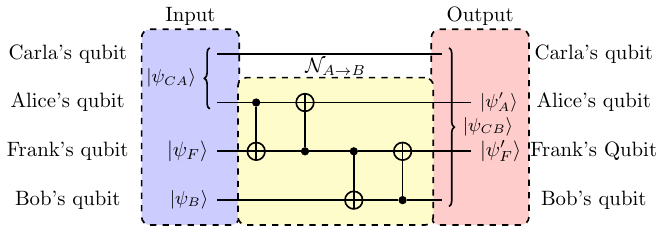}
\caption{Information is transferred from Alice to Bob through Frank as demonstrated with a quantum channel $\mathcal{N_{A\rightarrow B}}$, in this qubit-mediated QST channel.}\label{Swap_Intermediary_Qubit_2CNOTs}
\end{figure}

The diamond norm of the two channels is expressed as
\begin{equation}\label{diamond norm}
    \lVert \Xi_{A\rightarrow B}(\hat{\rho}_{A\varphi B}) - \mathcal{N}_{A\rightarrow B}(\hat{\rho}_{AFB}) \rVert_{\diamond} = sup\lVert \Xi_{A\rightarrow B}(\hat{\rho}_{A\varphi B}) - \mathcal{N}_{A\rightarrow B}(\hat{\rho}_{AFB})\rVert_1
\end{equation}
where we introduce our field-mediated QST channel as
\begin{equation}\label{quantum channel 2CNOTs with fields ch. 6}
    \Xi_{A\rightarrow B}(\hat{\rho}_{A\varphi B})= \Tr_{A\varphi}[\hat{U}^{QST}_{\varphi B}\hat{U}^{QST}_{A\varphi}( \hat{\rho}_{A,0} \otimes \kb{\psi_{\varphi,0}}{\psi_{\varphi,0}} \otimes \hat{\rho}_{B,0})\hat{U}^{QST \dagger}_{A\varphi} \hat{U}^{QST \dagger}_{\varphi B} ],
\end{equation}
the trace norm as $\lVert M \rVert_1 = \Tr\{|M|\}= \Tr\{\sqrt{MM^{\dagger}}\}$, and the supremum which maximizes over all possible arrangements of the initial states in the trace norm $\lVert \Xi_{A\rightarrow B}(\hat{\rho}_{A\varphi B}) - \mathcal{N}_{A\rightarrow B}(\hat{\rho}_{AFB})\rVert_1$. As we did in Sec.~\ref{Intermediate Qubit}, we are assuming that our initial states maximize this value for simplicity. For comparison we write down our qubit-mediated QST channel as
\begin{equation}\label{quantum channel 2CNOTs with qubits}
    \mathcal{N}_{A\rightarrow B}(\hat{\rho}_{AFB}) = \Tr_{AF}[\hat{U}_A \hat{U}_B( \hat{\rho}_{A,0}\otimes\kb{\psi_F}{\psi_F} \otimes \hat{\rho}_{B,0}) \hat{U}^{\dagger}_B \hat{U}^{\dagger}_A ]
\end{equation}
and diagrammatically we demonstrate Eq.~\eqref{quantum channel 2CNOTs with fields ch. 6} represented by Fig.~\ref{UDWQC Channel with starburst} and Eq.~\eqref{quantum channel 2CNOTs with qubits} by Fig.~\ref{Swap_Intermediary_Qubit_2CNOTs}. 

\begin{figure}
    \centering
    \begin{minipage}{.45\textwidth}
        \includegraphics[width=\textwidth]{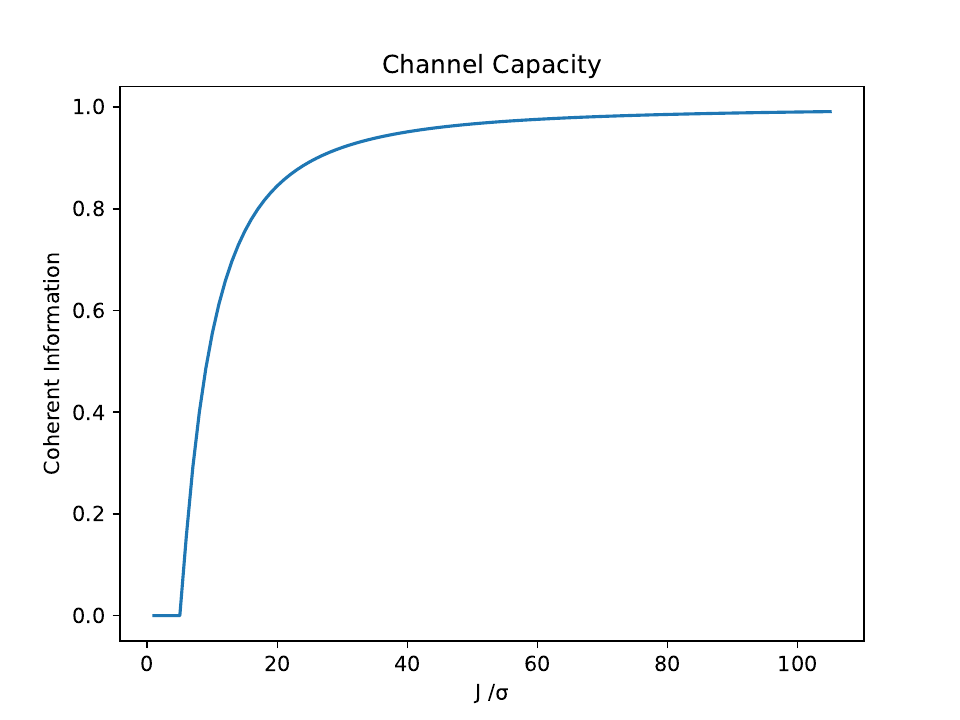}
        \caption{The channel capacity from Eq.~\eqref{quantum channel 2CNOTs with fields ch. 6}, With the initial state prepared in the $\ket{+_y}$ state. The significance of the ratio $J/\sigma$ will be briefly discussed in Sec.~\ref{Open Problems of Determining Noise}. For further investigations see Ref.~\cite{Simidzija2020Transmission}}
        \label{Channel_Capacity_graph}
    \end{minipage}
    \hspace{1cm}
    \begin{minipage}{.45\textwidth}
        \includegraphics[width=\textwidth]{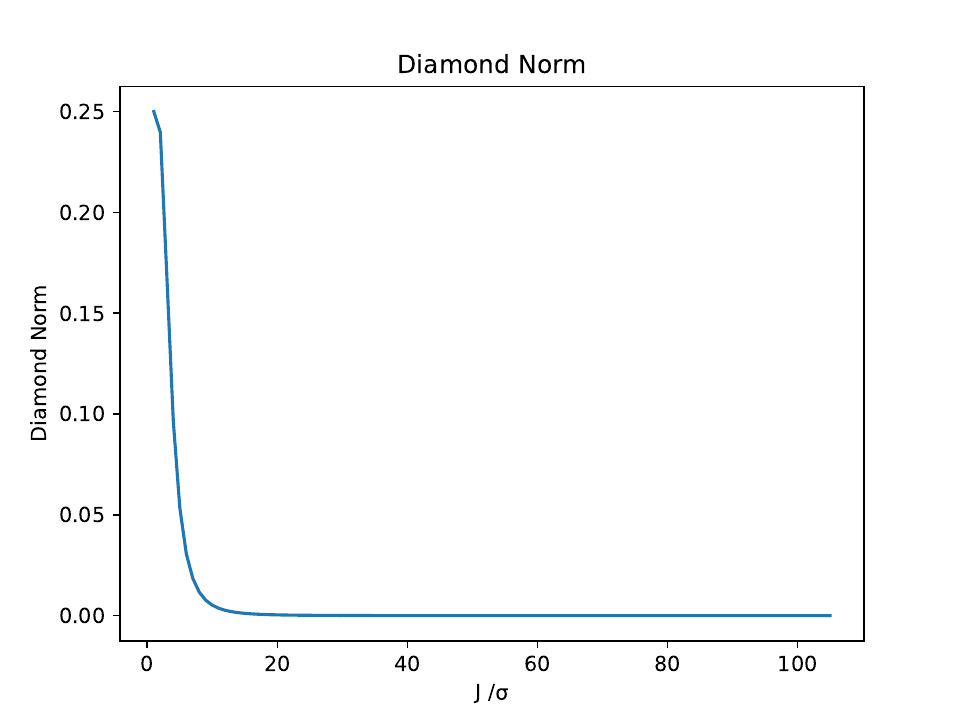}
        \caption{The diamond norm is a measure of ``idealness" of a channel. This particular graph compares the QST channel both qubit-mediated and field-mediated.}
        \label{diamond_norm_graph}
    \end{minipage}
\end{figure}

We can compare the diamond norm in Eq.~\eqref{diamond norm} to the channel capacity of Eq.~\eqref{quantum channel 2CNOTs with fields ch. 6}
\begin{equation}
     Q(\Xi_{A\rightarrow B}(\hat{\rho}_{A\varphi B}))= I_c(\mathbb{1}_C \rangle \Xi_{A\rightarrow B}(\hat{\rho}_{A\varphi B})) = S(\Xi_{A\rightarrow B}(\hat{\rho}_{A\varphi B})) - S(\mathbb{1}_C \otimes \Xi_{A\rightarrow B}(\hat{\rho}_{A\varphi B}))  
\end{equation}
which is numerically simulated with a Gaussian smearing function \cite{Simidzija2020Transmission,Aspling2023High} in Fig.~\ref{Channel_Capacity_graph} and Fig.~\ref{diamond_norm_graph} (The code for this simulation is available in Ref.~\cite{Aspling2023Information}). Preparation of initial states is different for Eqs.~\eqref{quantum channel 2CNOTs with qubits} and \eqref{quantum channel 2CNOTs with fields ch. 6}. This is a result of the different ways that the channels carry out the treatment of the information, and implies different methods are needed to carry out these two operations as formulated here.

\section{Universal Quantum Computing}

What we have shown in this letter, is that in the limit of strong coupling, two and three CNOT arrangements (QST and SWAP gates) can demonstrate equivalent processes between UDW field-mediated channels, and channels consisting of only qubits. With this established, we can now be a little more ambitious and seek universal quantum computing and communication. An important criteria of quantum computing, laid out by DiVincenzo \cite{DiVincenzo2000Physical} is that a system, capable of quantum computing, must have a set of universal quantum gates \cite{Bravyi2005Universal,DiVincenzo2000Universal}. With a little effort we can realize field-mediated CNOT gates as both qubit-mediated CNOT gates and single CNOT gates between two qubits. Furthermore, we can demonstrate these same processes for field-mediated single-qubit operations Hadamard, T-gate, and S-gate. Together with the CNOT, these single qubit gates form a universal set of quantum computing gates. 

\subsection{Forms of a single CNOT channel}\label{Forms of a single CNOT channel}

\begin{figure}
\centering
\begin{minipage}{.4\textwidth}
    \includegraphics[width=\textwidth]{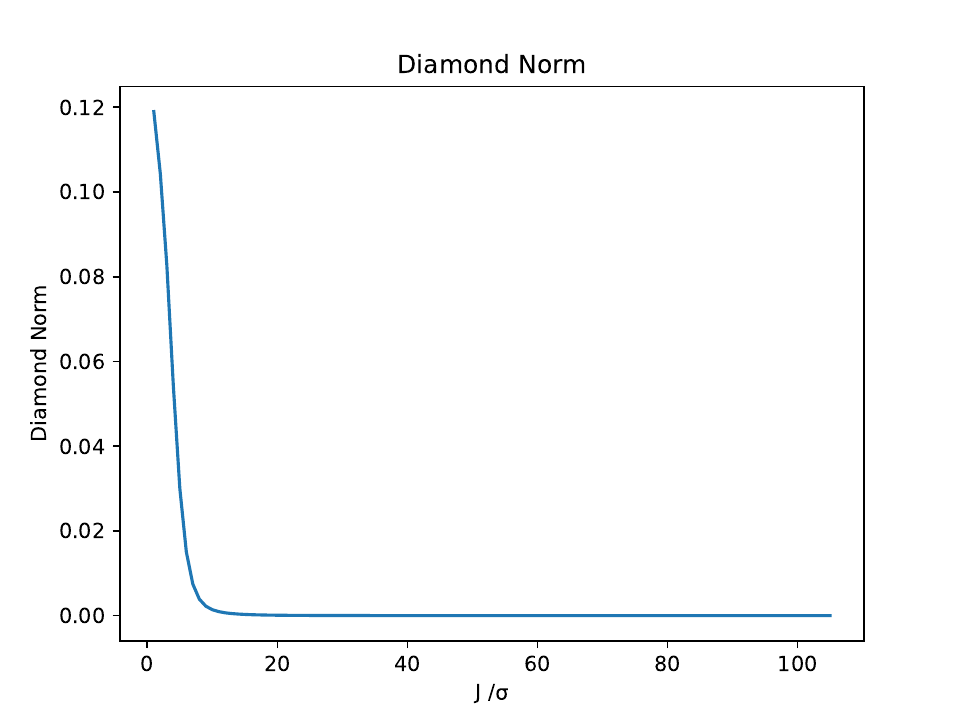}
        \caption{The diamond distance of the field-mediated CNOT channel equivalent to Eq.~\eqref{quantum channel 1CNOTs with fields} and the qubit-mediated CNOT channel.}
        \label{diamond_norm_graph_1CNOT}
\end{minipage}
\hspace{1cm}
\begin{minipage}{.4\textwidth}
    \includegraphics[width=\textwidth]{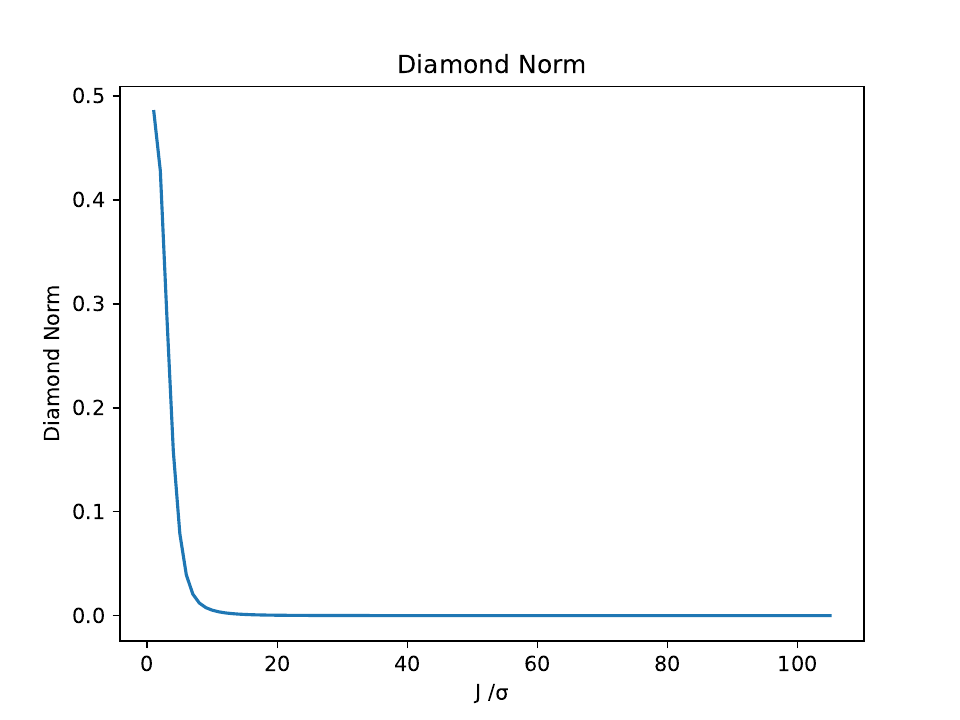}
        \caption{The diamond distance of a field-mediated CNOT gate equivalent to Eq.~\eqref{single CNOT with UDWs} and the standard CNOT between two qubits.}
        \label{diamond_norm_graph_1CNOT2qbits}
\end{minipage}
\end{figure}

As we have done previously, we introduce the field-mediated CNOT gate by directly comparing it to a Qubit-mediated CNOT channel. Qubit-mediated CNOT channels act firstly on two qubits with a single CNOT gate followed by a QST which transfers the state of the target to a final qubit as shown in Fig.~\ref{Swap_Intermediary_Qubit_2CNOTs}. To accomplish this with qubit-field gates we recall a single controlled unitary,
\begin{equation}\label{UDW qubit-mediated CNOT gate}
    \hat{U}^{\mathrm{Z\varphi}}_{\mathrm{QF}} = \sum_{\mu\in \pm} \hat{P}^{\mu}_z \otimes e^{i\mu\hat{\varphi}} = (\hat{P}^+_z \otimes e^{i\hat{\varphi}} + \hat{P}^-_z \otimes e^{-i\hat{\varphi}}).
\end{equation}
which demonstrates a channel equivalent to the qubit-mediated CNOT channel when written out as
\begin{equation}\label{quantum channel 1CNOTs with fields}
    \Xi_{A\rightarrow B}(\hat{\rho}_{A\varphi B}) = \Tr_{A\varphi}[\hat{U}^{\mathrm{X\Pi}}_{\varphi B}\hat{U}^{\mathrm{Z \varphi}}_{A\varphi}( \hat{\rho}_{A,0} \otimes \ket{\psi_{\varphi,0}}\bra{\psi_{\varphi,0}} \otimes \hat{\rho}_{B,0})\hat{U}^{\dagger\mathrm{Z\varphi}}_{A\varphi}\hat{U}^{\dagger\mathrm{X\Pi}}_{\varphi B}].
\end{equation}
where $\hat{U}^{\mathrm{X\Pi}}_{\varphi B}$ has taken the other form
\begin{equation}
    \hat{U}^{X\Pi}_{\mathrm{FQ}} = \sum_{\mu \in \pm} \hat{P}^{\mu}_x \otimes e^{i\mu\hat{\Pi}} = (\hat{P}^+_x \otimes e^{i\hat{\Pi}}+\hat{P}^-_x \otimes e^{-i\hat{\Pi}}).
\end{equation}
Here we have introduced the notation $\hat{U}^{\mathrm{Z\varphi}}_{\mathrm{QF}}$ where the superscripts label the qubit projector and field operator used, and the subscripts indicate the qubit and field being acted on. When initializing our receiving qubit (of the field mediated setup) in the $\ket{+_y}$ state, the diamond norm rapidly approaches zero as shown in  Fig.~\ref{diamond_norm_graph_1CNOT}.
\begin{figure}[ht]
\centering
\includegraphics{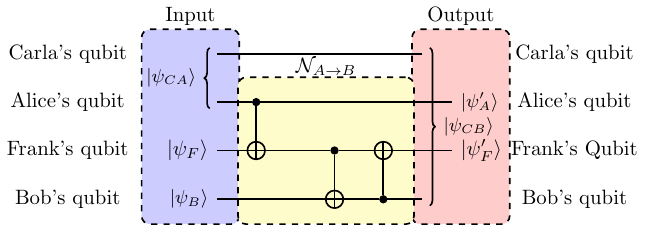}
\caption{A qubit-mediated CNOT gate has the equivalent output as a single CNOT gate between Alice and Bob.}\label{Swap_Intermediary_Qubit_CNOT}
\end{figure}

% The Hamiltonians representing these unitaries
% \begin{flalign}\label{UDW model Hamiltonians}
%     \hat{\mathrm{H}}_{\mathrm{A,int}}(t) &=  J_{A\varphi}\chi(t) \int_{\mathbb{R}} dx \ F(x) \hat{\mu}_A(t) \otimes \hat{\varphi}(x,t)\\
%     \hat{\mathrm{H}}_{\mathrm{B,int}}(t) &=  J_{B\varphi}\chi(t) \int_{\mathbb{R}} dx \ F(x) \hat{\mu}_B(t) \otimes \hat{\Pi}(x,t)\\
% \end{flalign}
% \michael{[Haven't we already introduced these Hamiltonians? Why introduce them again? Why now bring in Luttinger liquids and Kondo-like impurities? Best to just stick to qubit-field coupling hamiltians and not bring in extra baggage!]} are the electron density and electron current blocks of the Luttinger liquid Hamiltonian coupled with a Kondo-like impurity via the UDW interaction, a primary result in Ref.~\cite{Aspling2023Design}. These Hamiltonians indicate physical systems that can realize these gates. 

Surprisingly the field-mediated form of the two qubit CNOT gate like that of Eq.~\eqref{CNOTX} is bit more of a challenge. Regardless, we can realize this with the introduction of a new unitary \begin{flalign}
    \label{first 1/2 of single CNOT}
    \hat{U}^{Z\Pi X\varphi}_{\mathrm{QF}} =& e^{-i\hat{\Pi}} \sum_{\mu,\mu'\in \pm} \hat{P}^{\mu}_z\hat{P}^{\mu'}_x \otimes e^{i\mu\hat{\Pi}}e^{i\mu'\hat{\varphi}}\\ =& e^{-i\hat{\Pi}} (\hat{P}^+_z \otimes e^{i\hat{\Pi}} + \hat{P}^-_z \otimes e^{-i\hat{\Pi}})(\hat{P}^+_x \otimes e^{i\hat{\varphi}} + \hat{P}^-_x \otimes e^{-i\hat{\varphi}}) \nonumber
\end{flalign}
where the $e^{-i\hat{\Pi}}$ exists to absorb a left over phase. The channel therefore is expressed as 
\begin{equation}\label{single CNOT with UDWs}
    \Xi_{A\rightarrow B}(\hat{\rho}_{A\varphi B}) = \Tr_{A\varphi}[\hat{U}^{QST}_{\varphi B}\hat{U}^{Z\Pi X\varphi}_{A\varphi}( \hat{\rho}_{A,0} \otimes \kb{\psi_{\varphi,0}}{\psi_{\varphi,0}} \otimes \hat{\rho}_{B,0}) \hat{U}^{Z\Pi X\varphi \dagger}_{A\varphi} \hat{U}^{QST \dagger}_{\varphi B} ]
\end{equation}
and can be compared to a single CNOT gate with the diamond distance as shown in Fig.\ref{diamond_norm_graph_1CNOT2qbits}. The same process follows for Tofolli gates as well. 
\subsection{Field-Mediated Single qubit Operations}
A subset of the universal quantum gate arrangement is defined by single qubit operations. UDW 
\begin{wraptable}{r}{5.5cm}
\centering
    \begin{tabular}{|c|c|}
        \hline
        Input&Output\\
        \hline
        $\ket{0}$&$\frac{1}{\sqrt{2}}\left(\ket{0}+\ket{1}\right)$ \\
        $\ket{1}$&$\frac{1}{\sqrt{2}}\left(\ket{0}-\ket{1}\right)$ \\
        % Add more rows as needed
        \hline
    \end{tabular}
    \caption{Hadamard Gate creates a superpostion that depends on the initial state.}\label{Hadamard Gate}
\end{wraptable}
qubits describe the class of spin-qubits, well-known to be viable for all single qubit operations. However, we can demonstrate how one my construct an equivalent single qubit operator out of the controlled unitaries for our qubit-field interaction. Let's take the Hadamard operation with the truth Table~\ref{Hadamard Gate}.
It can be mimicked by defining a new unitary qubit-field operator
\begin{equation}\label{qubit-field Hadamard}
    \hat{U}^{\mathrm{H}}_{\mathrm{FQ}} = \sum_{\mu \in \pm} \hat{P}^{\mu}_x \otimes e^{-i\mu\hat{\varphi}}.
\end{equation}
Combining this with the unitary from Eq.~\eqref{UDW qubit-mediated CNOT gate}, into the quantum channel
\begin{equation}\label{Hadamard Channel with fields}
    \Xi^{\mathrm{H}}_{A\rightarrow A}(\hat{\rho}_{A\varphi}) = \Tr_{\varphi}[\hat{U}^{\mathrm{H}}_{\varphi A} \hat{U}^{\mathrm{Z\varphi}}_{A\varphi}( \hat{\rho}_{A,0} \otimes \kb{\psi_{\varphi,0}}{\psi_{\varphi,0}}) \hat{U}^{\mathrm{Z\varphi \dagger}}_{\varphi A}\hat{U}^{\mathrm{H}\dagger}_{\varphi A}]
\end{equation}
demonstrates the same computational effects of a single qubit Hadamard gate. Using the same logic we can show that there exist similar exact versions of single qubit S- and T-gates 
\begin{flalign}\label{qubit-field S and T}
    \hat{U}^{\mathrm{S}}_{\mathrm{QF}} &= \sum_{\mu\in \pm} \hat{P}^{\mu}_z \otimes (e^{(-\mu+1)i\hat{\Pi}}e^{i\hat{\varphi}})\\
    \hat{U}^{\mathrm{T}}_{\mathrm{QF}} &= \sum_{\mu\in \pm} \hat{P}^{\mu}_z \otimes (e^{\frac{-\mu+1}{2}i\hat{\Pi}}e^{i\hat{\varphi}}).
\end{flalign}
With theses gates we demonstrate, using the metric of diamond distance, that these gates are able to produce the same set of results as a universal set of logic gates.

\section{Open Problems of Determining Noise.}\label{Open Problems of Determining Noise}
As with any physical quantum computing system, one should expect a series of interactions that generate noise. It is worth noting a few known contributors to noise, while recognizing that there exists many unknown sources as well. For all intents and purposes, we treat the scalar fields in this article as free fields as well as made several other simplifying assumptions. However, the field observables here will exist in other blocks of the Hamiltonian $H = H_0 + H_1 + H_2...$. The time-evolution of the free field, often written in the zeroth order of the Hamiltonian $H_0$, is quadratic in nature and subsequently gives our coupling parameter units of $[L]^1$. In this article we evaluated low-energy and relatively close proximity interactions. For long-range (long-time) quantum communication this term we neglect becomes a highly relevant source of noise. 

Other terms that show up, in the bosonized form of the Tomonaga-Luttinger liquid for instance, are the forward scattering and back scattering interactions \cite{Senechal1999Introduction,Giamarchi2003Quantum}. In Ref.~\cite{Aspling2023Design}, the physical realizations of these gates are fermionic in nature and are only equivalent to scalar fields by identifying correlators between the two theories via bosonization\cite{shankar2017Bosonization}. The prescription methods in this paper, as well as in Ref.~\cite{Aspling2023Design}, allow for the evaluation of the forward scattering terms. The back scattering and Umklapp terms, introduce some complicated correlation functions that are difficult to evaluate. So Bosonization enables the identification of some contributions to the noise but also predicts others that may require significant computational resources to estimate. 

Beyond the mathematical noise, are the engineer restrictions. Quantum information is subject to Huygens' principle and with the law of no-cloning, is subject to rapid radial dispersion in solids \cite{Simidzija2020Transmission}. We have forgone these complications by remaining in (1+1) dimensions, which are experimentally realizable in Luttinger-liquids and optical fiber waveguides. Despite the progress of waveguide technology and topological protection supporting the Luttinger-liquids, information loss is still present and utilizing UDW quantum logic gates could provide numerical analysis to identify this noise and even protocols to limit it. 

\section{Conclusion}
We have shown that field-mediated CNOT gates, and processes that utilize multiple CNOT gates, can be realized through combinations of UDW unitary gates. This result is demonstrated through the channel capacity as well as the Diamond Norm between ideal gates and our field-mediated channels. Creating UDW channels needs to happen at the level of the field operators by constraining parameters. These constraints may or may not have a significant impact on the Hamiltonian and inhibit this from being implemented physically. Following the prescriptions outlined in this letter, one can construct sets of universal quantum computing gates, a key component in any quantum computing system but specifically needed in UDW applications. Accomplishing these field-mediated gates therefore opens up RQI to general applications in quantum Shannon theory, condensed matter physics, and provides insights into qubit-field transduction, a crucial part of long-range quantum communication.

\section{Acknowledgements}
The authors would like to thank Mark Wilde for helpful conversation and inspiration regarding the content of this paper. The authors would also like to thank Justin Kulp for many useful discussions.

\end{singlespace}
\bibliography{UDWQC_Logic_Gates.bib}
\end{document}